# Effect of Solution and Post-Mortem Time on Mechanical and Histological Properties of Liver during Cold Preservation

Mehmet Ayyildiz[1], Ranan Gulhan Aktas[2], Cagatay Basdogan[1]*

*[1]College of Engineering, Koc University, Istanbul*
*[2]School of Medicine, Koc University, Istanbul*

*Corresponding author: Email address: <u>cbasdogan@ku.edu.tr</u>.
Address: College of Engineering, Koc University, Istanbul, 34450, Turkey.
Tel.: +90 212 338 1721; Fax: +90 212 338 1548;

## Abstract

**BACKGROUND:** In liver transplantation, the donor and recipient are in different locations most of the time, and longer preservation periods are inevitable. Hence, the choice of the preservation solution and the duration of the preservation period are critical for the success of the transplant surgery.

**OBJECTIVE:** In this study, we examine the mechanical and histological properties of the bovine liver tissue stored in Lactated Ringer's (control), HTK and UW solutions as a function of preservation period.

**METHODS:** The mechanical experiments are conducted with a shear rheometer on cylindrical tissue samples extracted from 3 bovine livers and the change in viscoelastic material properties of the bovine liver is characterized using the fractional derivative Kelvin-Voigt Model. Also, the histological examinations are performed on the same liver samples under a light microscope.

**RESULTS:** The results show that the preservation solution and period have a significant effect on the mechanical and histological properties of the liver tissue. The storage and loss shear moduli, the number of the apoptotic cells, the collagen accumulation, and the sinusoidal dilatation increase, and the glycogen deposition decreases as the preservation period is longer.

**CONCLUSIONS:** Based on the statistical analyses, we observe that the liver tissue is preserved well in all three solutions for up to 11 h. After then, UW solution provides a better preservation up to 29 h. However, for preservation periods longer than 29 h, HTK is a more effective preservation solution based on the least amount of change in mechanical properties. On the other hand, the highest correlation between the mechanical and histological properties is observed for the liver samples preserved in UW solution.

# 1 Introduction

Liver is a vital organ responsible for many critical functions in the body. Currently, the function of liver cannot be compensated by any other ways and yet, it is not possible to treat a severe liver failure. Hence, a severely diseased liver must be removed from the patient and replaced with a whole or partial healthy liver. This surgical procedure is called liver transplantation. In this process, transplant liver, supplied from a living or a deceased donor, must be preserved and transported to a suitable recipient immediately. However, most of the time, the donor and the recipient are in different locations. Throughout the removal, preservation and transplantation, the stability of the interior environment of the liver is altered primarily due to the drop in its temperature (hypothermia) and insufficient supply of blood to its vessels (ischemia). As a result, the liver tissue inevitably experiences intracellular and extracellular changes such as cellular edema, cell death (apoptosis), depletion in cellular energy storage, hepatic sinusoidal dilatation, increase in connective tissue, acidosis, oxidation, and changes in ionic composition [1-4]. The longer the post-mortem time, the more damage develops in the liver. Various preservation solutions and methods are used to maintain the cell integrity, the integrity of the extracellular matrix (ECM), and pH balance of the liver tissue from the time of harvesting to transplantation. Flushing the liver with a preservation solution and storing it in the same solution at 0-4°C until the transplantation (cold storage) is the most widely used method for liver preservation [3].

The success of the liver transplantation primarily depends on the initial conditions of the transplant organ, type of the preservation solution and length of the preservation period between the procurement and the transplantation. However, there is still a dispute among the physicians on the choice of the preservation solution as well as on the extent of the preservation period. The first commonly used preservation solution for organ transplantation was EuroCollins (EC) solution, which was introduced in the 1970s [5]. The safe preservation period for storing a liver in this solution was four to eight hours. In the same decade, Marshall Solution was developed for liver and kidney preservation, which offered up to eight hours of safe liver preservation. In late 1980s,

University of Wisconsin (UW) solution was introduced for cold storage of the pancreas, liver, and kidney, and quickly become the "gold standard" for organ preservation in all over the world [6]. Despite the large success of the UW solution, other preservation solutions such as Histidine-Tryptophane-Ketoglutarate (HTK) and Celsior were also introduced to the market. Over the last two decades, HTK solution has become increasingly widespread especially in developing countries due to its lower cost and comparable outcomes with the UW solution. Today, UW and HTK stand as the most commonly used preservation solutions in the world for the liver transplantation. Although, these solutions vary in chemical composition, they all aim to delay the inevitable tissue damage. UW solution mimics an intracellular environment with high potassium and low sodium content. It is effective in protecting the integrity of the cell structure. In contrast, HTK is an extracellular type preservation solution with high sodium and low potassium concentrations and it is effective in preserving the integrity of the ECM structure. In addition, HTK solution has a lower viscosity (1.8 vs. 86.2 mPa.s), osmolarity (310 vs. 320 mOsm/L), and pH level (7.2 vs 7.4 pH) compared to UW solution [2, 7-9].

In spite of the extended information about the composition and rheological properties of UW and HTK solutions, it is still controversial which solution is better and how long an organ can be preserved safely in these solutions. Some investigators have reported that there are no major differences between the efficiencies of UW and HTK solutions in preservation. A study performed by Feng et al. [2] suggested that when the preservation period is short, there are no significant differences among UW, HTK, and Celsior solutions based on the incidence of biliary complications. Moench et al. [10] showed that there are no major differences between the UW and HTK solutions during in-situ and ex-situ perfusion of liver grafts at different pressure steps on pig models. Erhard et al. [11] examined the data of 60 patients who went through a liver transplantation.

They found no significant differences between the HTK and UW groups in terms of survival rate and the number of complications even when the cold ischemia period was more than 15 h. Testa et al. [12] evaluated the efficiencies of the UW and HTK solutions on adult-to-adult living donor liver transplantations. They perfused 30 grafts with either UW or HTK solutions. The authors reported that liver biochemistries, complications, and graft-patient survival rates of the two groups did not show any significant differences 13±7 months after the transplantation. However, they stated that using HTK solution was advantageous due to its lower cost. Also, it did not need to be flushed away before reperfusion of the graft.

Compared to the above results, some other studies in the literature reported that UW is a better solution with a longer preservation period. Straatsburg et al. [13] conducted a study to compare the effects of UW, HTK and Celsior solutions on cell death in 69 rat livers. The authors reported that UW and Celsior solutions were equally effective in preserving the rat liver cells for 0-16 h, but they were best preserved in UW solution after 24 h. Den Toom et al. [14] investigated the effects of UW and Euro-Collins (EC) solutions on the morphology and metabolic capacity of rat livers for 18 to 42 h of preservation. They reported that UW solution provided better morphology and metabolic capacity compared to EC solution after 18 h of preservation. They also reported that the differences between UW and EC solutions were more strengthened after 42 h of preservation. A study conducted by Todo et al. [15] analyzed 185 cadaveric liver homografts preserved for 4 to 24 h in UW solution and 180 grafts preserved 3 to 9.5 h in EC solution. The livers stored in UW solution had a higher survival rate and lower primary non-function rate, although the preservation duration of the liver in UW solution was about two times longer than that of the livers in EC solution. In a recent study, Stewart et al. [16] analyzed the data of 21626 patients, who underwent deceased donor kidney transplantation between the years 2004 and 2008, to determine the effects of UW and HTK solutions on the delayed graft function and death-censored graft survival (i.e. graft loss without death). The authors found that HTK preservation showed 20% increase in the death-censored graft loss compared to UW preservation. Other studies [17-19] also reported that UW is better than HTK in kidney preservation when the preservation time is more than 24 h.

In this study, we investigate the changes in mechanical and histological properties of the liver tissue simultaneously as a function of preservation period to provide better insight into the effectiveness of UW and HTK solutions during liver preservation, and also the upper limit of safe preservation period in these solutions. During the cold ischemic preservation, the histological properties of the liver tissue in micro scale change with increasing post-mortem times and these changes lead alterations in gross mechanical properties at macro scale. However, the relation between the histological and gross mechanical properties during cold preservation has not been investigated at all. Understanding this relation may lead to the development of new protocols for the evaluation of the tissue injury during preservation and hence the suitability of the donor liver for transplantation.

In order to investigate the changes in mechanical properties of the liver tissue as a function of preservation solution and period, we conduct dynamic shear loading (DSL) and quasi-static shear loading (QSSL) experiments with a rheometer on bovine liver samples preserved in Lactate Ringer, UW and HTK solutions for 5 h, 11 h, 17 h, 29 h, 41 h, and 53 h. In order to quantify the changes in mechanical properties, a fractional derivative viscoelastic model is fit to the experimental data and the corresponding model parameters are reported for each preservation period and solution. In order to investigate the changes in histological properties under a light microscope, small tissue samples taken from the same livers are stained with different chemicals to count the number of apoptotic cells, to measure the collagen accumulation, sinusoidal dilatation, and glycogen level in the liver cells. Finally, the effects of preservation solution and period on the mechanical and histological properties are investigated by two-way ANOVA first and then the statistical differences between the individual groups are further analyzed by Bonferroni corrected paired t-tests. Also, the correlations between the mechanical and histological measurements are investigated by Spearman's Rank-Order correlation method.

## 2   Material and Methods

### 2.1   Preparation of Tissue Samples

Thin cylindrical tissue samples having a flat surface are required for rheological shear experiments. However, preparing such samples is challenging due to the viscoelastic nature of the liver tissue.  In order to obtain tissue samples from a bovine liver in desired thickness and diameter, a tissue slicing and sampling apparatus has been developed[1] (Figure 1a). The commercial devices (e.g. microtome) used in histology and pathology laboratories of hospitals are designed to obtain very small and thin samples. Moreover, before obtaining these samples, soft tissue is fixated to reduce morphological distortions and damage, which makes it easier to obtain very small and thin samples. However, the tissue samples used in our experiments have to be fresh and cylindrical in shape, with a diameter of 20-50 mm and thickness of 2-3 mm as suggested in [20] for rheological experiments. Furthermore, since we investigate the effect of 3 different preservation solutions and 6 different preservation periods on the material properties of the liver, we need to obtain several samples from each liver. Obtaining several cylindrical samples from liver tissue in desired diameter and thickness is a challenging task since soft tissue does not keep its form under any force, including the one due to gravity. However, it is highly important to apply force to immobilize the tissue during cutting to obtain a slice with a uniform thickness. For this purpose, we use a vacuum pump and immobilize the soft tissue by suction. The assembly order of our design is shown in Figure 1b and its part list is given in Table 1.

---

[1] The tissue slicing and sampling apparatus presented in this article was inspired by a similar apparatus developed by Juergen Braun at Charité-Universitätsmedizin Berlin, Germany.

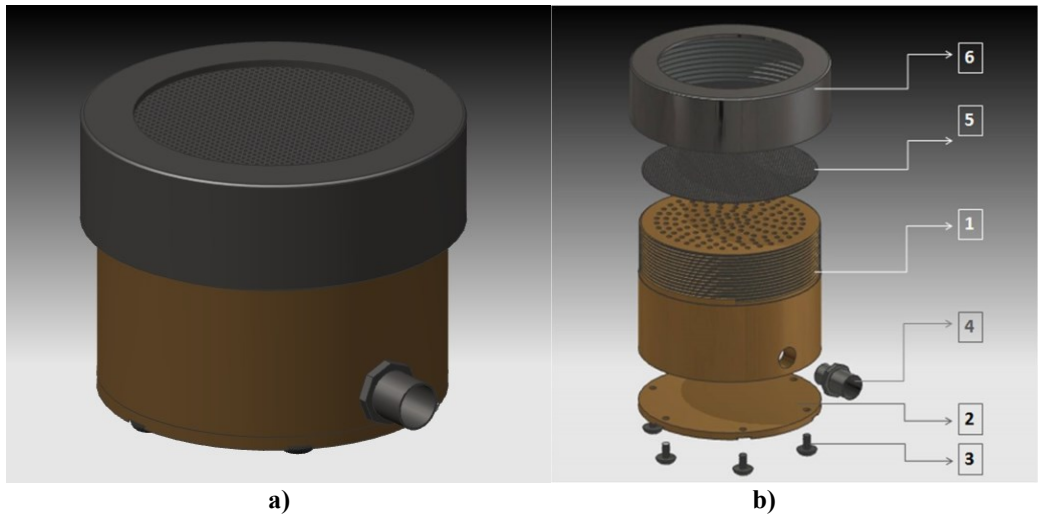

**Figure 1** The tissue slicing and cutting apparatus: **a)** solid model and, **b)** assembly order.

**Table 1** The part list of our tissue slicing and cutting apparatus

| Part No | Part Name | Number of Units |
|---------|-----------|-----------------|
| 1 | Vacuum Chamber | 1 |
| 2 | Lower Plate | 1 |
| 3 | M4 x 8 Screw | 6 |
| 4 | Connecting Adapter | 1 |
| 5 | Perforated Thin Plate | 1 |
| 6 | Cylindrical Nut | 1 |

In this study, three fresh bovine livers are obtained from a slaughterhouse in three consecutive weeks and cold preserved in Lactated Ringer's solutions at +4ºC immediately after the harvesting. The livers are transferred from the slaughterhouse to our laboratory in 4 h. In order to prepare soft tissue samples, the following steps are performed; first, the gap between the top surface of the cylindrical nut and the perforated plate of the tissue slicing and cutting apparatus is adjusted to 2 mm (Figure 2a). Then, a chunk of liver excluding the Glisson's capsule and visible blood vessels are cut and placed on the top surface of the apparatus (Figure 2b). The vacuum pump is turned on, the liver tissue is sucked through the holes of the perforated plate into the cylindrical nut, and immobilized. Using a sharp slicing blade, small oscillatory movements are applied to the immobilized tissue to remove the excessive layer at the top. Hence, a tissue slice with a thickness of 2 mm and a diameter of 80 mm is obtained from the liver tissue as shown in Figure 2c. The vacuum pump is turned off and the cylindrical nut is unthreaded. Then, the tissue slice is taken out of the apparatus and put on a flat surface (Figure 2d). Finally, a cylindrical blade (Figure 2e) is used to obtain 4 cylindrical samples having a diameter of 25 mm and a thickness of $2 \pm 0.5$ mm from each slice (Figure 2f).

A total of 54 cylindrical tissue samples (3 preservation solutions × 6 preservation periods × 3 repetitions) from the left lobe (lobus hepatis sinister) is obtained from each liver. The samples are divided into 3 groups and stored in 3 different solutions at +4°C: UW and HTK for the preservation purposes and Lactated Ringer's solution for the control purposes. The preparation of the cylindrical tissue samples took another hour following the four-hour transportation period. Consequently, the tissue samples placed in Lactated Ringer's, HTK and UW solutions are subjected to mechanical measurements at the 5th h, 11th h, 17th h, 29th h, 41st h, and 53rd h of the cold ischemic preservation.

Note that the tissue slicing process takes no more than one minute (Figure 2b-2c) and does not cause any significant de-hydration. Nicolle and Palierne (2010) [21] reported that the de-hydration process is fully reversible when the samples are re-hydrated. In our experiments, the samples were put into a preservation solution (Ringer, HTK, UW) right after the preparation of the tissue samples. The holes of the perforated plate are sufficiently small not to cause any damage on the tissue slices during suction. All the slices are cut in the same orientation (parallel to the perforated thin plate shown in Figure 1) in order to reduce undesired measurement errors due to anisotropy. Each sample is tested within 20 minutes to ensure that samples are not suffered from dehydration during the rheological experiments as suggested in [21].

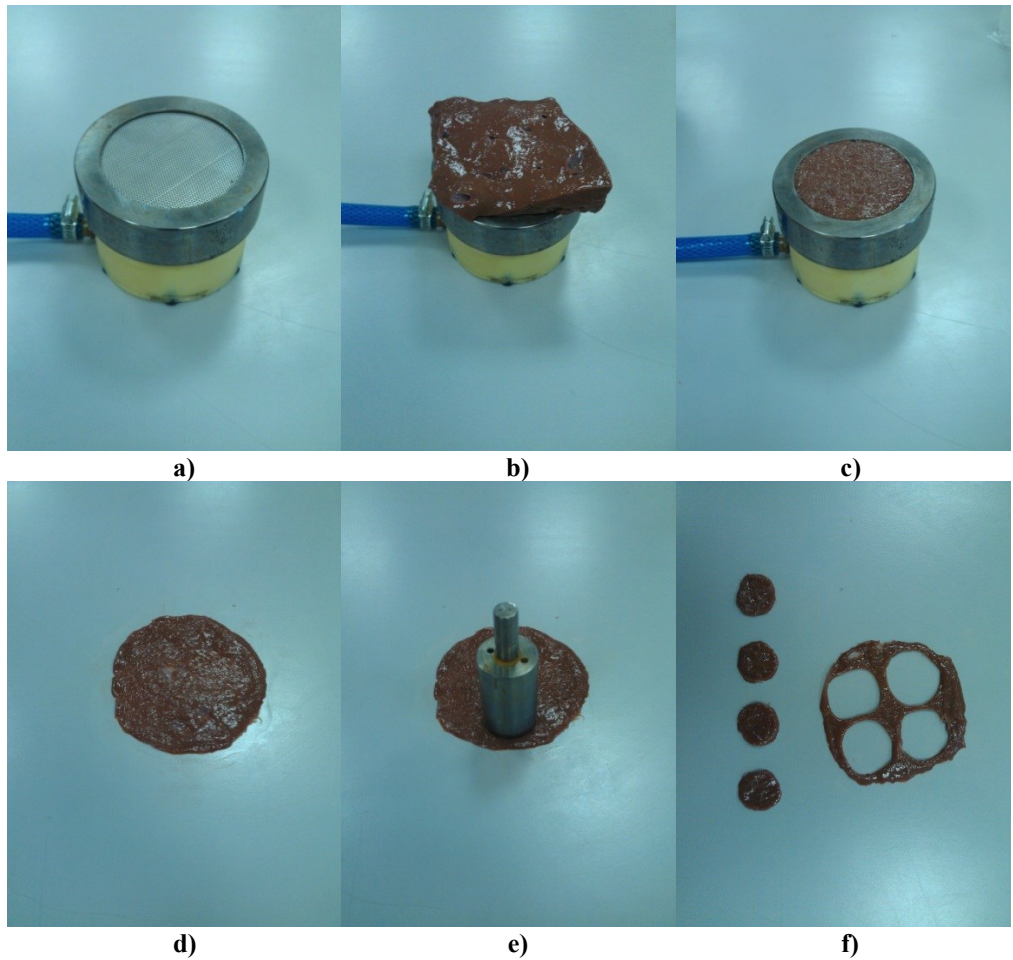

**Figure 2** The steps of the tissue slicing and sampling: **a)** the gap between the cylindrical nut and the perforated plate is adjusted to 2 mm, **b)** the liver is immobilized on the top surface of the apparatus by suction, **c)** the excess liver is removed using the sharp slicing blade and the tissue slice with a diameter of 80 mm is left in the apparatus after the liver was cut, **d)** the vacuum pump is turned off, the cylindrical nut is unthreaded, the slice is taken out of the apparatus and placed on a flat surface, **e)** a sharp cylindrical blade is used to obtain samples from the slice, and then **f)** four samples, each with a diameter of 25 mm, were obtained from each slice.

For the histological analyses, small tissue blocks having a thickness of 3 mm are obtained from the parenchyma of the left lobe of each liver and put into Ringer, HTK and UW solutions at the 5[th] preservation period. These small samples are fixed in 10% neutral buffered formalin at 11 h, 17 h, 29 h, 41 h, and 53 h preservation periods for 24 h at room temperature and dehydrated by bathing in a graded series of mixtures of ethanol and water. Also, it should be emphasized that for the control purposes, the same procedure is followed for the small samples preserved in Ringer solution at the 5[th] preservation period. Afterwards, hydrophobic clearing agent (xylene) is used to remove the alcohol and also, an infiltration agent (paraffin wax) is used to replace xylene. Then, the samples in paraffin wax are heated in the oven at 60°C for 2 h. Finally, these samples are sectioned at 7-10 µm in thickness using a microtome and stained with different kits to analyze the histological properties under a light microscope.

## 2.2 Examination of the Mechanical Properties

We perform DSL and QSSL experiments on the cylindrical tissue samples preserved in Lactated Ringer's, HTK and UW solutions for 5 h, 11 h, 17 h, 29 h, 41 h, and 53 h using a rheometer (Anton Paar, MCR 102). The temperature during the experiments is controlled at T = 4º C by the Peltier module (P-PTD200/56/AIR) of the rheometer. Each cylindrical liver sample is put between the concentric parallel plates of the rheometer as shown in Figure 3a. In order to determine the height of the sample, the upper plate of the rheometer is commanded to move towards the sample with a very slow speed (50 µm/s). When the normal force is reached to 0.5 N, the movement of the upper plate is stopped and the gap between the upper and lower plates is measured and taken as the height (H) of the sample (Figure 3b). Afterwards, a pre-strain of 5% is applied to the samples in the normal direction to maintain full contact between the tissue sample and the parallel plates of the rheometer as suggested in the literature [22-24]. In addition, sandpapers having a grain size of P80 are attached to the parallel plates of the rheometer using a double-sided tape (3M - 9473PC) to ensure that no slippage occurs between the plates and the sample during the experiments as recommended in the literature [23, 25-27]. Then, pre-conditioning is applied to the sample to reduce the variations among the measurements as recommended by the other groups in the literature [23, 24, 28]. For this purpose, the sample is oscillated at 1 Hz frequency at 0.5% torsional shear strain for 50 cycles before the DSL and QSSL experiments.

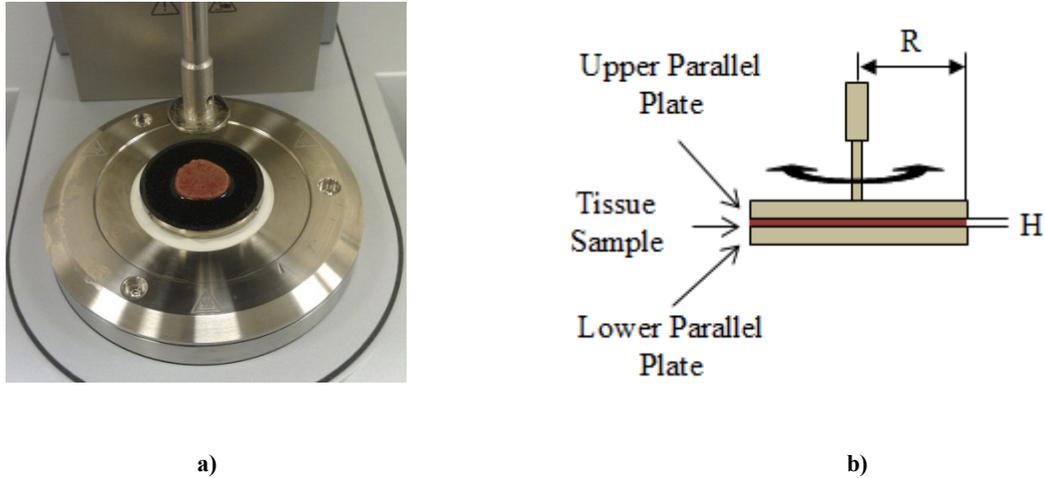

a)                                                                                                          b)

**Figure 3 a)** A cylindrical tissue sample held between two parallel plates of the rheometer, **b)** a schematic showing the plates and the sample (R: radius of the parallel plates; H: height of the sample)

### 2.2.1 Dynamic Shear Loading (DSL) Experiments

The most common method for characterizing the frequency-dependent viscoelastic material properties of soft tissues is the frequency sweep test. In this experiment, the upper plate of the rheometer oscillates at varying frequencies ($\omega$) with an amplitude of $\gamma_A$ to introduce a sinusoidal shear strain, $\gamma^*$ (Equation 1).

$$\gamma^* = \gamma_A \, e^{i\omega t} = \gamma_A \, \sin(\omega t)$$

Equation 1

During the oscillation, the reaction torque is measured to calculate the complex shear stress, $\tau^*$ (Equation 2),

$$\tau^* = \tau_A \, e^{i(\omega t + \delta)} = \tau_A \, \sin(\omega t + \delta)$$

<div align="right">Equation 2</div>

where, $\tau_A$ is the amplitude of the shear stress and $\delta$ is the phase angle. Finally, the complex shear ($G^*$), storage ($G_S$) and loss ($G_L$) moduli are calculated using the Equation 3-4.

$$G^* = \frac{\tau^*}{\gamma^*}$$

<div align="right">Equation 3</div>

$$G^* = G_S + iG_L$$

<div align="right">Equation 4</div>

In our DSL experiments, the frequency range of the oscillation and the amplitude of the shear strain are set to $\omega=0.1$-20 Hz and $\gamma_A=0.5\%$, respectively. The frequency range of the oscillations is determined based on the measurement bandwidth of our rheometer and the recommendations given in [29].

### 2.2.2 Quasi-Static Shear Loading (QSSL) Experiments

Each tissue sample is loaded up to 15% torsional shear strain ($\gamma_A$) with a very slow angular velocity ($0.0005 \text{ s}^{-1}$). The reaction torque is measured and converted to the torsional shear stress ($\tau_A$). Tangent shear modulus, the slope of the torsional stress-strain curve, is calculated at shear strains of $\gamma_A = 0.1\%$, 5% and 15%.

### 2.2.3 Modeling

A Kelvin-Voigt fractional derivative model (KVFDM) is used to estimate the optimum viscoelastic material coefficients of liver tissue by minimizing the error between the experimental data and the corresponding values generated by the model using a nonlinear least square fit algorithm. Since we measure the torsional shear stress, our KVFDM consists of a torsional spring and a torsional spring-pot element connected in parallel. It is capable of representing both -time (relaxation, creep) and -frequency (dynamic oscillation) dependent behavior of liver tissue. This model has been successfully utilized by other researchers in the literature for representing the viscoelastic behavior of soft tissues [22, 27, 30]. The constitutive equations for KVFDM in time and frequency domains are given in Equation 5 and 6. Here, V [Pa s$^\propto$] is a unique quasi-property of the material, which shows the magnitude of the material response, $D^\propto[\gamma_A(t)]$ is the fractional derivative of shear strain with the order of $\propto$, where $0 < \propto < 1$. Since, V [Pa s$^\propto$] contains a non-integer power of time, it is not a true material property, but a quasi-property [31, 32]. Quasi-properties are used to compactly describe the viscoelastic relaxation spectrum via only few parameters. Also, the spring-pot element can be described as the interpolation between a spring ($\propto$ = 0, V = E [Pa]) and a dash-pot ($\propto$ = 1, V = $\eta$ [Pa s]). The meaning of the spring-pot element is better appreciated in vectorial representation as suggested in [32].

If both sides of the Equation 6 are divided by the shear strain $\gamma_A(\omega)$, the complex shear modulus, G*($\omega$), is obtained (Equation 7), which can be separated into real (storage modulus, $G_S$) and imaginary (loss modulus, $G_L$) parts (Equation 8). Then, the tangent of the phase angle between stress and strain, $\delta$, is calculated using Equation 9.

$$\tau_A(t) = G\,\gamma_A(t) + V\,D^\propto[\gamma_A(t)]$$

Equation 5

$$\tau_A(\omega) = G\,\gamma_A(\omega) + V\,(i\omega)^\propto\,\gamma_A(\omega)$$

Equation 6

$$G^*(\omega) = G + V\,(i\omega)^\propto$$

Equation 7

$$G^*(\omega) = \left[G + V\cos\left(\frac{\pi\,\propto}{2}\right)\omega^\propto\right] + i\left[V\sin\left(\frac{\pi\,\propto}{2}\right)\omega^\propto\right]$$

Equation 8

$$\tan(\delta) = \frac{V\sin\left(\frac{\pi\,\propto}{2}\right)\omega^\propto}{G + V\cos\left(\frac{\pi\,\propto}{2}\right)\omega^\propto}$$

Equation 9

In our approach, we measure the linear shear modulus of the liver samples (G at $\gamma_A = 0.1\%$) during the QSSL experiments. We use this value in the KVFDM (Equation 8) to estimate the viscoelastic material coefficients of the spring-pot element (V and ∝) for each preservation solution and period using the data recorded for the DSL experiments. This approach enables us to estimate a unique set of material coefficients. Otherwise, the curve-fit is not unique.

## 2.3  Examination of the Histological Properties

The tissue sections are colored with four different stains for histological examination: (1) Apop-Tag Plus Peroxidase kit (Intergen S7101, Merck Millipore Inc.), also known as TUNEL (Terminal deoxynucleotidyl transferase dUTP nick end labeling) technique, is used to examine the hepatocytes undergoing apoptosis, (2) Masson`s Trichrome stain (Masson-Goldner Staining kit, Merck Millipore Inc.) is used to detect the changes in connective tissue, (3) Hematoxylene (Hematoxylene solution modified to Gill III, Merck) & Eosin (Eosin Y solution 0.5 % alcoholic, Merck Millipore Inc.) (H&E) is used to detect structural changes in sinusoids, and (4) Periodic-Acid Schiff stain (PAS Staining kit, Merck) is used to detect the changes in glycogen level in the liver cells.

All treated sections are examined under a light microscope (Axio Imager, Carl Zeiss Inc.). The microscopic images, captured from ten different areas on each section, are examined at 100X magnification by Axio Vision image analysis software package (Carl Zeiss Inc.). Four software scripts are developed (1) to count the nuclei of apoptotic cells labeled by Apop-Tag Plus Peroxidase kit, (2) to measure the collagen accumulation on each section stained with Masson`s trichrome, (3) to calculate the area of sinusoids on each section stained with H&E, and (4) to measure the glycogen level in the cells on each section stained with PAS.

## 2.4    Correlation between the Mechanical and Histological Properties

One of the goals of this study is to provide an insight into the relation between the mechanical and histological properties of the liver tissue. For this purpose, the changes in gross mechanical properties of the liver tissues are correlated with the histological properties in micro scale using the Spearman's Rank-Order Correlation method. Before the analyses, the histological data collected for each preservation period from ten different areas of each tissue section is averaged for three animals (the average of ten histological measurements × 5 preservation periods × 3 livers = 15 data points for each histological property). Also, the mechanical data collected from three tissue samples for each preservation period is averaged for three animals (the average of three mechanical measurements × 5 preservation periods × 3 livers = 15 data points for each mechanical property). In the correlation analyses, the significance level is chosen as p=0.05 and the strength of the association between two data sets is evaluated based on the correlation coefficient ($r_s$) as very strong (VS, $1.0 \geq |r_s| \geq 0.8$), strong (S, $0.8 > |r_s| \geq 0.6$), moderate (M, $0.6 > |r_s| \geq 0.4$), weak (W, $0.4 > |r_s| \geq 0.2$) or very weak (VW, $0.2 > |r_s| \geq 0$). Additionally, two-way ANOVA test is used to investigate the effect of the preservation period, solution, and their interaction on the mechanical properties of the liver. Then, Bonferroni corrected paired t-tests are performed to further assess the statistical differences between the individual groups.

# 3    Results

## 3.1    Mechanical Properties

### 3.1.1    Dynamic Shear Loading (DSL) Experiments

The storage and loss moduli of the liver samples preserved in Ringer, HTK and UW solutions for 5 – 53 h are plotted as a function of torsional shear strain ($\omega$=20 Hz) in Figure 4. The amplitude of the shear strain ($\gamma_A$) used in the frequency sweep experiments is selected within the linear regime based on the results of the amplitude sweep experiments. These experiments are performed at $\omega$=20 Hz for the shear strain varying between $\gamma_A$ =0.1-2%. The linear viscoelastic regime is determined as $\gamma_{LVR}$=1% based on 10% deviation of either the storage or the loss modulus from the initial value (the value at $\gamma_A$ =0.1%).

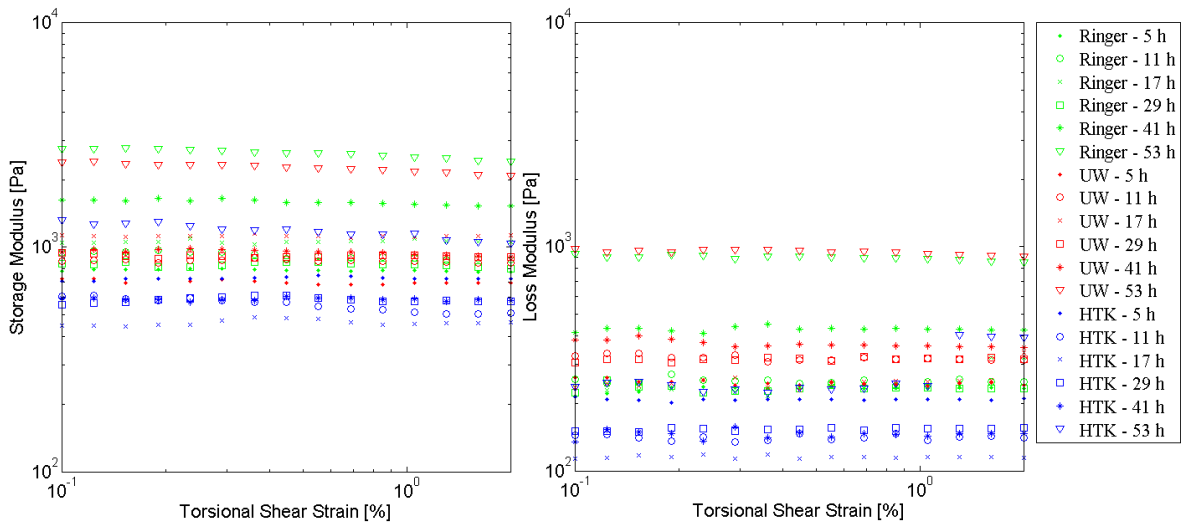

**Figure 4 a)** The storage and **b)** loss moduli of bovine liver samples as a function of preservation period, solution and torsional shear strain.

The average storage and loss moduli of the liver samples preserved in Ringer, HTK and UW solutions for 5 – 53 h are plotted as a function of all frequencies tested ($\omega$=0.1-20 Hz) in Figure 5 and more specifically for the frequencies of ω=0.1 Hz, 1 Hz and 10 Hz in Figure 6. Also, the average and standard deviation values of storage and loss moduli for ω=0.1 Hz, 1 Hz and 10 Hz are tabulated in Table 2.

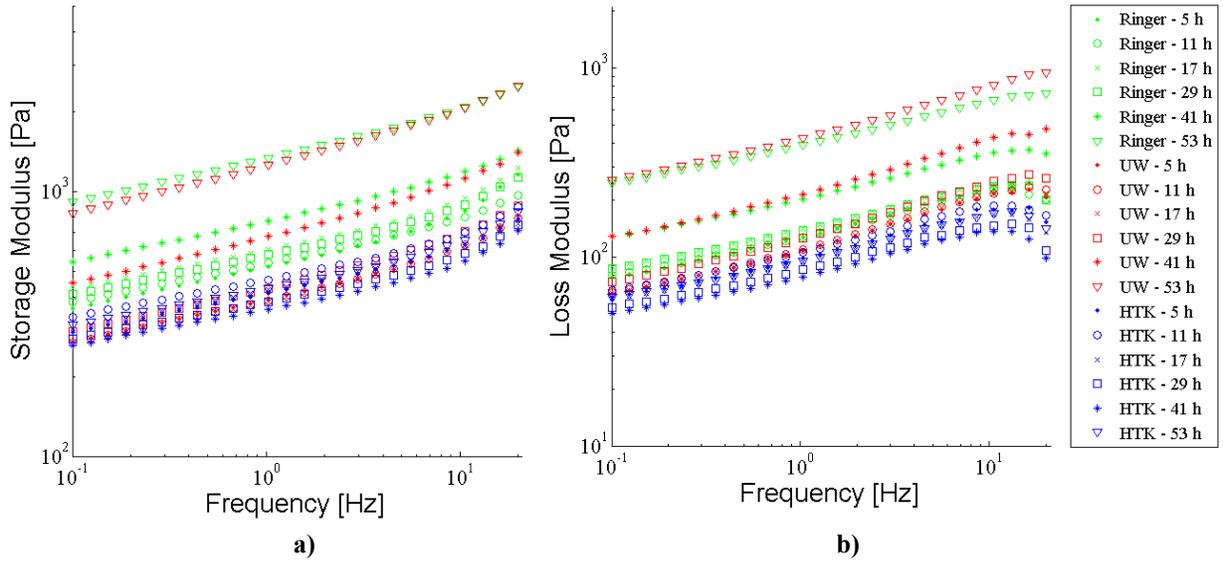

**a)**                                                        **b)**

**Figure 5 a)** The storage and **b)** loss moduli of bovine liver samples as a function of preservation period, solution and oscillation frequency.

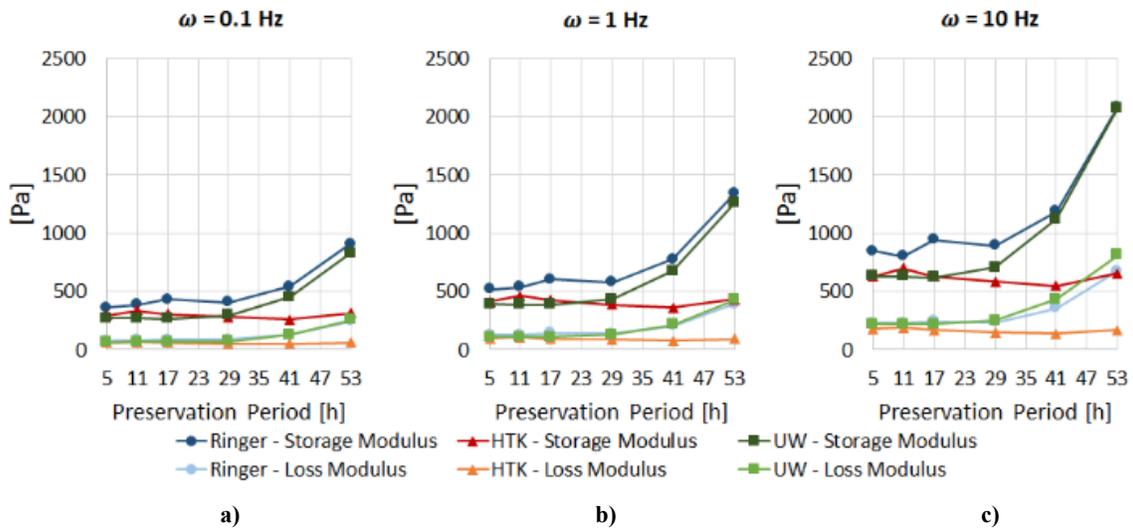

**a)**                                  **b)**                                  **c)**

**Figure 6** The storage and loss moduli of the liver tissue as a function of preservation solution and period for the frequency of stimulation at **a)** 0.1 Hz, **b)** 1 Hz, and **c)** 10 Hz.

**Table 2** The average and standard deviation of the storage and loss moduli of bovine liver samples at 0.1 Hz, 1 Hz and 10 Hz with respect to the preservation solution and period.

| Preservation Period [h] | Frequency [Hz] | Ringer | | HTK | | UW | |
|---|---|---|---|---|---|---|---|
| | | Storage Modulus [Pa] | Loss Modulus [Pa] | Storage Modulus [Pa] | Loss Modulus [Pa] | Storage Modulus [Pa] | Loss Modulus [Pa] |
| 5 | 0.1 | 361±31 | 78±7 | 295±47 | 61±6 | 271±45 | 64±9 |
| | 1 | 520±40 | 128±12 | 414±60 | 98±12 | 392±61 | 109±16 |
| | 10 | 850±132 | 230±26 | 632±84 | 179±25 | 632±71 | 217±34 |
| 11 | 0.1 | 385±22 | 81±4 | 338±26 | 67±3 | 269±19 | 65±4 |
| | 1 | 539±24 | 128±8 | 465±30 | 106±6 | 389±22 | 110±6 |
| | 10 | 801±40 | 228±16 | 701±37 | 187±16 | 630±46 | 217±13 |
| 17 | 0.1 | 431±25 | 89±9 | 305±20 | 61±6 | 266±12 | 65±4 |
| | 1 | 605±32 | 142±11 | 426±34 | 95±6 | 386±12 | 109±12 |
| | 10 | 944±30 | 245±17 | 628±44 | 168±8 | 622±47 | 217±38 |
| 29 | 0.1 | 409±5 | 87±2 | 278±38 | 54±4 | 296±90 | 74±18 |
| | 1 | 578±4 | 138±3 | 385±45 | 86±8 | 433±122 | 126±31 |
| | 10 | 894±26 | 235±6 | 585±51 | 147±19 | 707±182 | 252±61 |
| 41 | 0.1 | 544±148 | 128±34 | 262±63 | 51±9 | 452±83 | 128±28 |
| | 1 | 776±207 | 203±52 | 360±82 | 79±16 | 676±134 | 216±46 |
| | 10 | 1189±305 | 358±88 | 547±123 | 137±29 | 1122±216 | 430±85 |
| 53 | 0.1 | 911±318 | 247±109 | 314±91 | 61±12 | 829±32 | 256±7 |
| | 1 | 1338±496 | 390±174 | 432±110 | 95±17 | 1261±37 | 425±7 |
| | 10 | 2084±773 | 675±306 | 658±142 | 171±36 | 2073±55 | 815±20 |

The storage and loss moduli of the tissue samples increase as a function of frequency. The storage and loss moduli of the tissue samples stored in Ringer and UW solutions increase as the preservation period is longer. However, the storage and loss moduli of the samples stored in HTK solution show relatively small change. For example, at low frequency ($\omega$=0.1 Hz), the storage modulus of the samples stored in Ringer, HTK and UW solutions are between $361 - 911$ Pa, 262-338 Pa and 266-829 Pa, respectively. Also, the loss modulus of the same samples are between 78-247 Pa, 51-67 Pa and 64-256 Pa, respectively. When the frequency is increased to 10 Hz, the storage modulus of the samples stored in Ringer, HTK and UW solutions varies between $801 - 2084$ Pa, 547 - 701 Pa and 622-2073 Pa, respectively. In addition, the loss modulus of the same samples are between 228-675 Pa, 137-187 Pa and 217-815 Pa, respectively. In order to investigate the impact of the preservation solution and period on the storage and loss moduli of the liver tissue, two-way ANOVA is performed. The results show that the preservation period, solution and their interaction have significant effects on the storage and loss moduli of the livers (p<0.05). Furthermore, paired t-tests are performed to evaluate the statistical difference between the measurements performed at different preservation periods. The results suggest that the storage modulus of the liver samples preserved in Ringer, HTK and UW solutions change significantly at the $11^{th}$ h (p <0.05), $11^{th}$ h (p<0.001) and $29^{th}$ h (p<0.001), respectively. On the other hand, the loss modulus of the same samples preserved in Ringer, HTK and UW solutions change significantly (p<0.001) at the $17^{th}$ h, $29^{th}$ h and $29^{th}$ h, respectively. In particular, the storage moduli of the samples preserved in Ringer and UW solutions increase almost 2-folds at the $41^{st}$ h (p<0.001) and almost 4-folds at the $53^{rd}$ h (p<0.001) compared to the measurements obtained at the $5^{th}$ h. However, the storage and loss moduli values of the samples preserved in HTK solution show relatively small changes (1.1 folds, p<0.001) throughout the whole testing period.

### 3.1.2    Quasi-Static Shear Loading (QSSL) Experiments

The average torsional shear stress and tangent shear modulus of the liver samples preserved in Ringer, HTK and UW solutions for 5 – 53 h are given as a function of torsional shear strain in Figure 7. Moreover, the average and standard deviation of the tangent shear modulus at shear strain of 0.1%, 5% and 15% are tabulated in Table 3. As shown in Figure 7, the torsional shear stress increases while the tangent shear modulus decreases as the torsional shear strain is increased. Also, it is observed that the samples stored in the Ringer solution have the highest tangent shear modulus while the samples stored in the HTK solution have the lowest tangent shear modulus. Similar to the results of the frequency sweep experiments, the torsional shear stress and tangent shear modulus values of the samples preserved in Ringer and UW solution increase as the preservation period is increased. However, the changes in the torsional shear stress and the tangent shear modulus of the samples stored in HTK are relatively small. At small torsional shear strain (0.1%), the tangent shear modulus of the samples stored in Ringer, HTK and UW solutions varies between 265 – 631 Pa, 224-261 Pa and 190-443 Pa, respectively. At large torsional shear strain (15%), the tangent shear modulus of the samples stored in Ringer, HTK and UW solutions varies between 123 – 242 Pa, 103 - 130 Pa and 84-166 Pa, respectively. Moreover, two-way ANOVA tests demonstrate that the preservation period, solution and their interaction have significant effects on the torsional shear stress and tangent shear modulus of the livers (p<0.05). Additionally, the paired t-tests show that the torsional shear stress and tangent shear modulus of the liver samples preserved in Ringer, HTK and UW solutions change significantly at 11th h (p<0.001). At torsional shear strain of 0.1%, the tangent shear modulus of the samples stored in Ringer and UW solutions for 53 h is 2.4-folds (p<0.001) and 2.2-folds (p<0.001) higher than those of the values for 5 h, respectively. However, at torsional shear strain of 0.1%, the tangent shear modulus increases only 1.2-folds (p<0.001) for the samples stored in HTK solution throughout the whole testing period.

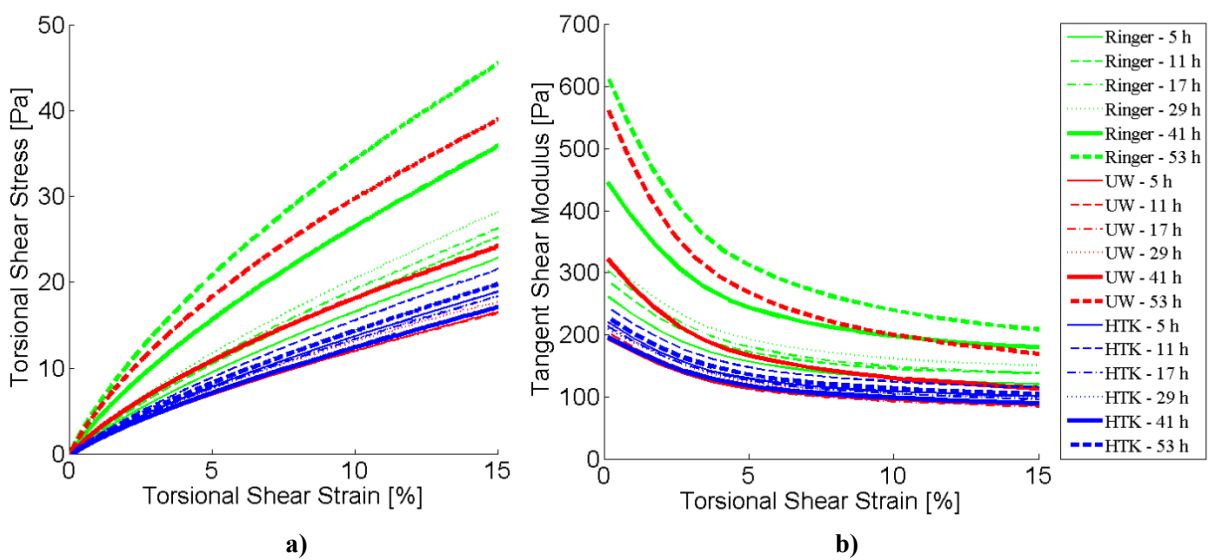

**a)**                                **b)**

**Figure 7 a)** The torsional shear stress and **b)** tangent shear modulus of the bovine liver samples as a function of preservation period, solution and oscillation frequency.

**Table 3** The average and standard deviation of the tangent shear modulus of bovine liver samples at torsional shear strain of 0.1%, 5% and 15%.

| Preservation Period [h] | Torsional Shear Strain [%] | Ringer | HTK | UW |
|---|---|---|---|---|
| | | Tangent Shear Modulus [Pa] | Tangent Shear Modulus [Pa] | Tangent Shear Modulus [Pa] |
| 5 | 0.1 | 265±22 | 224±38 | 198±26 |
| | 5 | 156±24 | 129±30 | 114±22 |
| | 15 | 123±22 | 103±26 | 88±19 |
| 11 | 0.1 | 292±16 | 261±23 | 214±26 |
| | 5 | 174±18 | 157±14 | 123±18 |
| | 15 | 141±18 | 130±12 | 95±14 |
| 17 | 0.1 | 322±22 | 241±31 | 190±19 |
| | 5 | 195±18 | 142±20 | 108±9 |
| | 15 | 161±21 | 115±15 | 84±7 |
| 29 | 0.1 | 328±7 | 229±20 | 219±51 |
| | 5 | 196±16 | 135±17 | 124±36 |
| | 15 | 160±15 | 107±13 | 94±28 |
| 41 | 0.1 | 467±93 | 235±25 | 322±28 |
| | 5 | 265±68 | 139±17 | 175±19 |
| | 15 | 204±59 | 111±15 | 124±16 |
| 53 | 0.1 | 631±40 | 235±58 | 443±64 |
| | 5 | 340±28 | 136±44 | 244±39 |
| | 15 | 242±16 | 108±40 | 166±21 |

### 3.1.3 Modeling

We fit the KVFDM to the complex shear modulus data collected from the DSL experiments as a function of preservation solution and period (Figure 8a). In our approach, the tangent shear modulus of the bovine liver at $\gamma_A$=0.1% shear strain (Table 3) is used as the G parameter in the KVFDM to obtain a unique set of viscoelastic parameters. The estimated viscoelastic coefficients (V and $\propto$) and the coefficient of determination ($R^2$) for $G^*$, $G_S$, $G_L$ are tabulated in Table 4. The vectorial representations of the curve-fit results for the liver samples preserved for 5 and 53 h are given in Figure 8b. In this figure, the x- and y- axes represent the energy storage and dissipation capabilities of the liver samples, respectively. It is observed that as the preservation period increases from 5 h to 53 h, the magnitude of the complex shear modulus of the liver samples preserved in Ringer and UW solutions increase about 3-folds. However, those preserved in HTK does not change significantly. Figure 9 presents the change in the viscoelastic material coefficients of liver tissue as a function of preservation solution. Moreover, Figure 10 shows the phase angle as a function of frequency, preservation solution and period. The phase angle of the liver samples preserved in Ringer and UW solutions increase as a function of frequency and preservation solution, but those preserved in HTK solution show a small change.

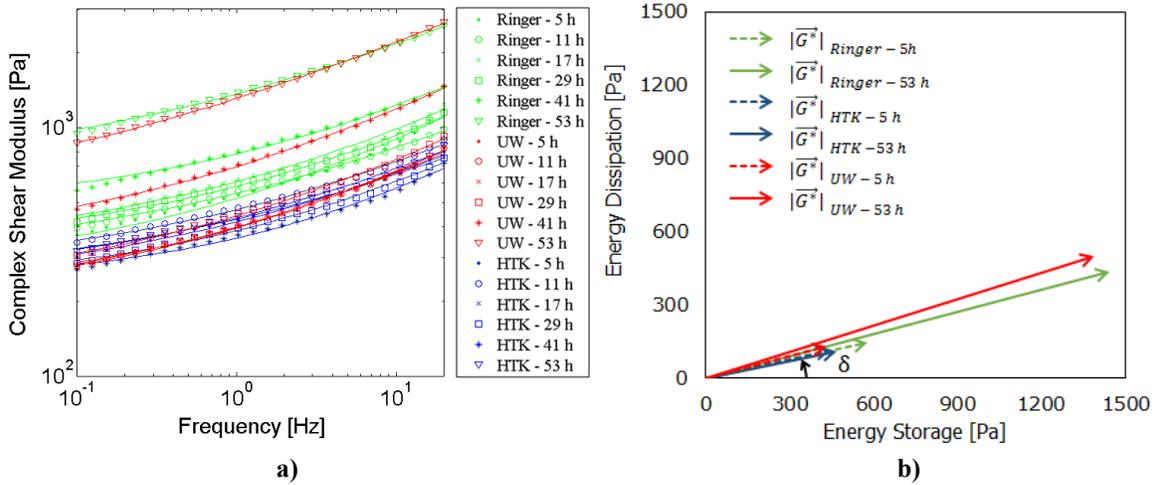

**a)**           **b)**

**Figure 8 a)** The complex shear modulus of the bovine liver samples as a function of preservation period, solution and frequency. KVFDM is fit to the experimental data (solid lines). **b)** The vectorial representation of curve-fit results for the liver samples preserved for 5 and 53 h. The y-axis represents the energy dissipation capability, while the x-axis shows the energy storage capability of the liver tissue.

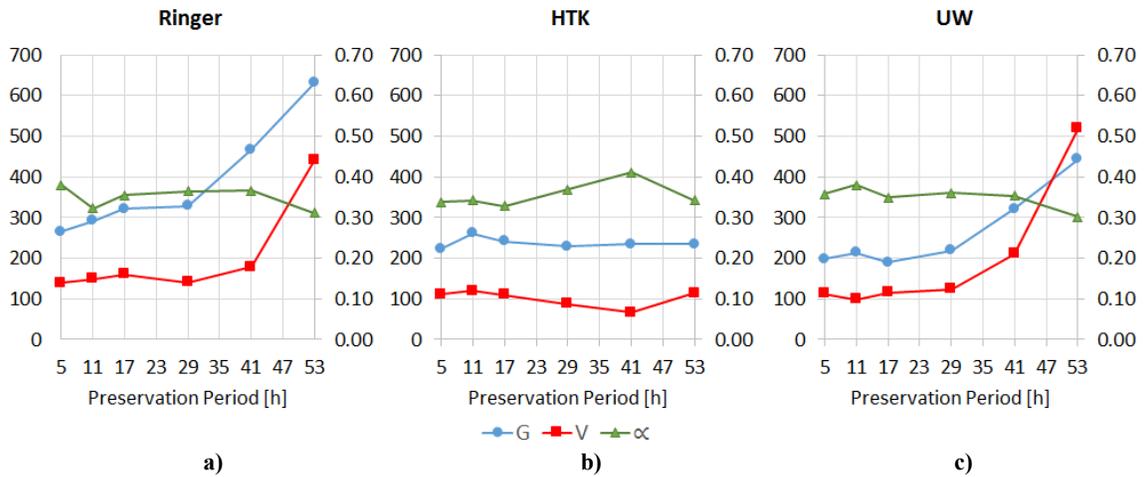

**a)**           **b)**           **c)**

**Figure 9** The viscoelastic material coefficients of liver tissue preserved in **a)** Ringer, **b)** HTK, and **c)** UW solutions. The y-axis on the left hand side is the scale for the G and V variables. The y-axis on the right hand side is the scale for the ∝ variable.

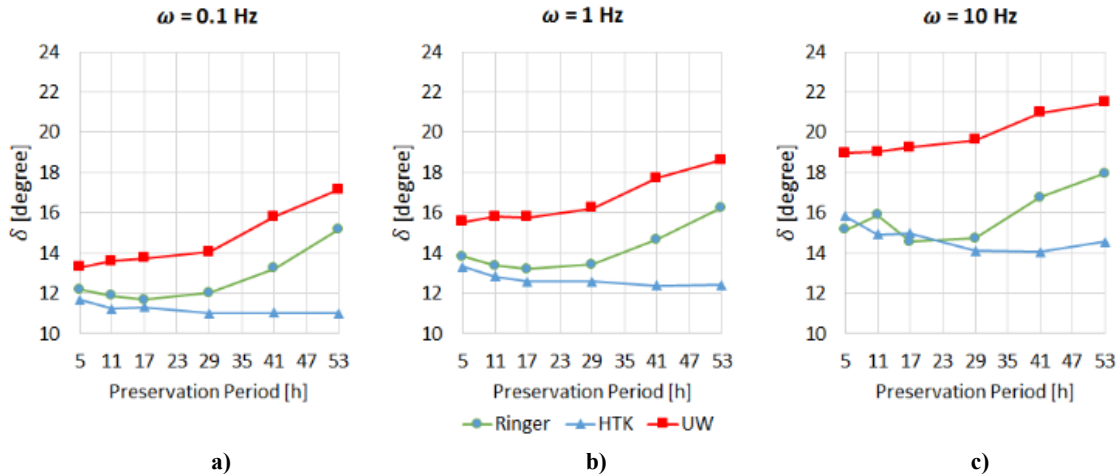

**Figure 10** The phase angle, δ, as a function of preservation solution and period for the frequency of **a)** 0.1 Hz, **b)** 1 Hz, and **c)** 10 Hz.

**Table 4** The viscoelastic material coefficients of liver tissue as a function of preservation period and solution. ($R^2$ is the coefficient of determination).

| Preservation Period [h] | Viscoelastic Parameter | Ringer | HTK | UW |
|---|---|---|---|---|
| | V [Pa s$^\propto$] | 139 | 112 | 112 |
| 5 | $\propto$ | 0.38 | 0.34 | 0.36 |
| | R$^2$ (for G*, G$_S$ G$_L$) | 0.99, 0.98, 0.95 | 1.00, 0.99, 0.93 | 1.00, 0.99, 0.97 |
| | V [Pa s$^\propto$] | 149 | 119 | 99 |
| 11 | $\propto$ | 0.32 | 0.34 | 0.38 |
| | R$^2$ (for G*, G$_S$ G$_L$) | 1.00, 1.00, 0.93 | 1.00, 0.99, 0.92 | 1.00, 0.99, 0.98 |
| | V [Pa s$^\propto$] | 161 | 110 | 115 |
| 17 | $\propto$ | 0.36 | 0.33 | 0.35 |
| | R$^2$ (for G*, G$_S$ G$_L$) | 1.00, 0.99, 0.91 | 1.00, 1.00, 0.89 | 1.00, 0.99, 0.98 |
| | V [Pa s$^\propto$] | 141 | 87 | 124 |
| 29 | $\propto$ | 0.36 | 0.37 | 0.36 |
| | R$^2$ (for G*, G$_S$ G$_L$) | 1.00, 0.99, 0.90 | 1.00, 0.99, 0.83 | 1.00, 1.00, 0.98 |
| | V [Pa s$^\propto$] | 179 | 67 | 210 |
| 41 | $\propto$ | 0.37 | 0.41 | 0.35 |
| | R$^2$ (for G*, G$_S$ G$_L$) | 1.00, 1.00, 0.97 | 1.00, 0.99, 0.79 | 1.00, 1.00, 0.99 |
| | V [Pa s$^\propto$] | 441 | 114 | 519 |
| 53 | $\propto$ | 0.31 | 0.34 | 0.30 |
| | R$^2$ (for G*, G$_S$ G$_L$) | 1.00, 1.00, 0.99 | 1.00, 0.99, 0.90 | 1.00, 1.00, 1.00 |

## 3.2 Histological Properties

We analyze the changes in the number of the apoptotic cells, the connective tissue accumulation, the sinusoidal dilatation, and the level of glycogen of the tissue blocks fixed at the $11^{th}$, $17^{th}$, $29^{th}$, $41^{st}$ and $53^{rd}$ h of cold ischemic preservation (10 liver sections × 3 preservation solutions × 5 preservation periods = a total of 150 sections are analyzed). Also, the histological properties of the tissue blocks stored in Ringer solution and fixed at the $5^{th}$ h preservation period are examined for the control purposes (10 liver sections × 1 preservation solutions × 1 preservation period = a total of 10 sections are analyzed). The results of the histological analyses are summarized in Table 5.

**Table 5** The histological properties of bovine liver (average of 3 animals).

| Preservation Period [h] | Apoptotic Cell [count] | | | Connective Tissue [%] | | | Sinusoidal Dilatation [%] | | | Glycogen Level [%] | | |
|---|---|---|---|---|---|---|---|---|---|---|---|---|
| | Ringer | HTK | UW | Ringer | HTK | UW | Ringer | HTK | UW | Ringer | HTK | UW |
| 5 | 13±6 | | | 6±3 | | | 28±13 | | | 36±8 | | |
| 11 | 26±14 | 42±9 | 6±5 | 7±3 | 7±4 | 7±5 | 31±11 | 31±7 | 22±6 | 25±10 | 25±10 | 34±13 |
| 17 | 37±11 | 43±9 | 8±6 | 10±5 | 8±4 | 11±6 | 30±6 | 36±12 | 28±7 | 14±9 | 14±9 | 36±15 |
| 29 | 48±13 | 59±16 | 19±13 | 14±7 | 8±5 | 15±6 | 37±13 | 35±11 | 32±8 | 10±8 | 17±10 | 20±10 |
| 41 | 44±12 | 58±17 | 31±20 | 15±8 | 8±4 | 15±8 | 31±5 | 42±10 | 30±6 | 8±8 | 12±10 | 12±8 |
| 53 | 44±9 | 57±13 | 34±21 | 16±6 | 13±8 | 17±8 | 39±7 | 45±10 | 43±7 | 3±2 | 6±5 | 9±8 |

The apoptotic cells in each tissue section are identified by Apop Tag Plus Peroxidase kit. The brown/dark brown cells on the micrographs are marked and counted to determine the number of apoptotic cells. The change in the number of apoptotic cells as a function of preservation solution and period is given in Figure 11a. The results show that the number of apoptotic cells increase as a function of preservation period. For the samples stored in Ringer and HTK solutions, the apoptotic cell count increase significantly after the $11^{th}$ h, while those preserved in UW solution increases after the $29^{th}$ h compared to the control group ($p<0.05$). The samples stored in HTK solution have the highest number of apoptotic cells, while the ones stored in UW solution have the lowest number of apoptotic cells at the end of the testing period (53 h).

The accumulation of connective tissue is analyzed by the Masson`s trichrome stain. The connective tissue is colored as green in the micrographs and its area is measured. The change in the accumulated connective tissue as a function of preservation solution and period is plotted in Figure 11b. For the samples stored in Ringer and UW solutions, the connective tissue accumulates significantly after the $17^{th}$ h compared to the samples in the control group (p<0.05). Nevertheless, for the samples preserved in HTK solution, the connective tissue accumulates significantly only after $53^{rd}$ h compared to the samples in control group (p<0.05). The results suggest that the connective tissue accumulates as a function of preservation period regardless of the preservation solution. At the end of the testing period (53 h), the highest and the lowest accumulation is observed in the samples stored in UW and HTK solutions, relatively.

The tissue sections are stained with H&E to quantify the degree of sinusoidal dilatation. The borders of the sinusoids are outlined with red lines in the micrographs and the areas enclosed by these borders are measured. The change in the sinusoidal dilatation of the bovine liver as a function of preservation solution and period is plotted in Figure 11c. The results suggest that the sinusoidal dilatation increases as a function of preservation period. For the samples preserved in Ringer, HTK and UW solutions, the sinusoidal dilatation increases significantly after the $29^{th}$ h, $17^{th}$ h, and $53^{rd}$ h, compared to the samples in the control group, respectively (p<0.05). The biggest and smallest dilatation is detected in the samples preserved in HTK and Ringer solutions at the $53^{rd}$ h, respectively.

The glycogen level in the cells of the bovine liver is examined with the PAS stain. The blue-magenta colored areas in the tissue sections are detected and measured by the image analysis software. The change in the glycogen level is plotted as a function of preservation solution and period in Figure 11d. It is clear that the glycogen level decreases as a function of preservation period. Compared to the samples in the control group, the glycogen level in the tissue samples preserved in Ringer and HTK solutions decreases significantly after the 11th h (p<0.05) and after the 29th h in the samples preserved in UW solution (p<0.05). When the glycogen levels are inspected at the 53rd h for each solution, it is observed that the samples stored in UW solution has the highest level while those preserved in Ringer solution has the lowest level.

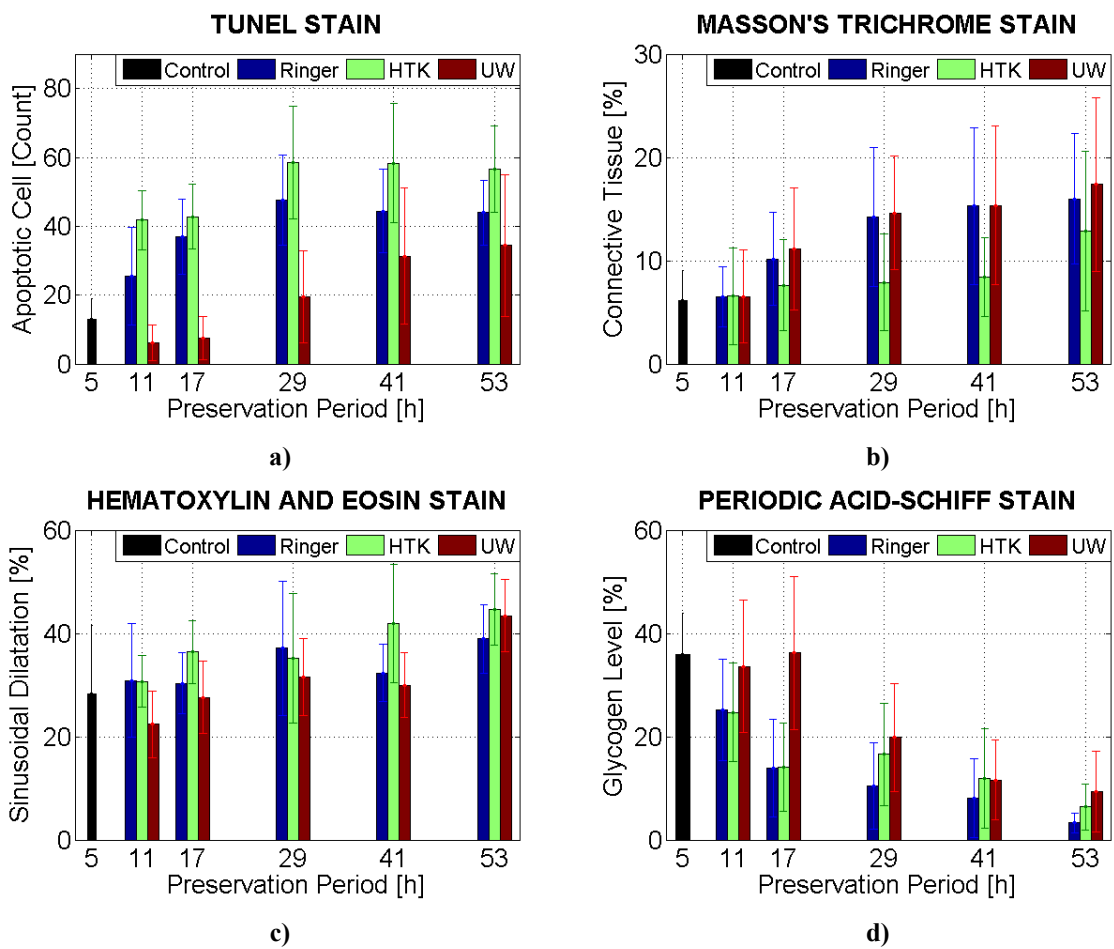

**Figure 11** The change in the **a)** apoptotic cell count, **b)** connective tissue, **c)** sinusoidal dilatation, and **d)** glycogen level as a function of preservation solution and period (average of 3 animals).

## 3.3 Correlation Between the Mechanical and Histological Properties

The correlation coefficients and the strengths of the correlations between the mechanical and histological properties of the liver samples preserved for 11 h – 53 h in Ringer, HTK and UW solutions are tabulated in Table 6, respectively. The results of the correlation analyses show that the apoptotic cell count is strongly correlated with the amount of connective tissue, but not correlated with the other parameters for the samples stored in Ringer and HTK solutions. Also, the apoptotic cell count is moderately correlated with the dynamic mechanical properties at low frequencies, strongly correlated with the dynamic mechanical properties at high frequencies, and moderately correlated with the quasi-static mechanical properties for the samples stored in UW solution. The connective tissue accumulation is strongly/very strongly correlated with the quasi-static and dynamic mechanical properties for the samples preserved in Ringer and UW solutions. But, there is no correlation between these parameters for the samples stored in HTK solution. The sinusoidal dilatation is moderately/strongly correlated (negative) with the glycogen level for the samples stored in all preservation solutions. The sinusoidal dilatation is not correlated with the quasi-static and dynamic mechanical properties of the samples stored in Ringer solution. However, it is correlated with the loss modulus at low frequency and tangent shear modulus at high strain for the samples stored in HTK solution. Also, it is moderately/strongly correlated with the mechanical properties stored in UW solution. The glycogen level is moderately/very strongly and strong/very strongly correlated (negative) with the mechanical properties for the samples stored in Ringer and UW solutions, respectively. Nevertheless, it is not correlated with the mechanical properties of the samples stored in HTK solution. Generally speaking, the quasi-static and dynamic mechanical properties are very strongly correlated with each other for the samples stored in all solutions.

**Table 6** The correlation coefficients and the strength of correlation (1.0≥|rs|≥0.8: 'Very Strong (VS)'; 0.8>|rs|≥0.6: 'Strong (S)'; 0.6>|rs|≥0.4: 'Moderate (M)'; 0.4>|rs|≥0.2: 'Weak (W)', 0.2>|rs|≥0: 'Very Weak (VW)'; p<0.05: 'SI: Statistically Insignificant') between the mechanical and histological properties of bovine liver samples preserved for 5 h – 53 h in (Ringer, HTK, and UW) solutions.

| Parameter | | Apoptotic Cell [Count] | Connective Tissue [%] | Sinusoidal Dilatation [%] | Glycogen Level [%] | Storage Modulus [Pa] | | Loss Modulus [Pa] | | Tangent Shear Modulus [Pa] | |
|---|---|---|---|---|---|---|---|---|---|---|---|
| | | | | | | ω=0.1 Hz | ω=10 Hz | ω=0.1 Hz | ω=10 Hz | γ=0.1% | γ=15% |
| Apoptotic Cell [Count] | | - | 0.75,0.62,SI | SI,SI,SI | SI,SI,SI | SI,SI,0.52 | SI,SI,0.61 | SI,SI,0.58 | SI,SI,0.62 | SI,SI,0.52 | SI,SI,0.57 |
| Connective Tissue [%] | | S,S,SI | - | SI,0.58,0.77 | SI,SI,-0.88 | 0.61,SI,0.69 | 0.60,SI,0.75 | 0.75,SI,0.73 | 0.61,SI,0.81 | 0.72,SI,0.83 | 0.59,SI,0.80 |
| Sinusoidal Dilatation [%] | | SI,SI,SI | SI,M,S | - | -0.6,-0.6,-0.69 | SI,SI,0.51 | SI,SI,0.56 | SI,-0.53,0.52 | SI,SI,0.61 | SI,SI,0.69 | SI,-0.61,0.65 |
| Glycogen Level [%] | | SI,SI,SI | SI,SI,VS | S,M,S | - | -0.77,SI,-0.74 | -0.88,SI,-0.81 | -0.76,SI,-0.80 | -0.64,SI,-0.84 | -0.58,SI,-0.84 | -0.60,SI,-0.86 |
| Storage Modulus [Pa] | ω=0.1 Hz | SI,SI,M | S,SI,S | SI,SI,M | S,SI,S | - | 0.88,0.71,0.93 | 0.93,0.91,0.95 | 0.89,0.91,0.93 | 0.77,SI,0.91 | 0.90,0.74,0.88 |
| | ω=10 Hz | SI,SI,S | S,SI,S | SI,SI,M | VS,SI,VS | VS,S,VS | - | 0.9,0.72,0.99 | 0.81,0.71,0.99 | 0.73,SI,0.91 | 0.76,0.52,0.88 |
| Loss Modulus [Pa] | ω=0.1 Hz | SI,SI,M | S,SI,S | SI,M,M | S,SI,S | VS,VS,VS | VS,S,VS | - | 0.95,0.95,0.98 | 0.80,SI,0.91 | 0.83,0.71,0.90 |
| | ω=10 Hz | SI,SI,S | S,SI,VS | SI,SI,S | S,SI,VS | VS,VS,VS | VS,S,VS | VS,VS,VS | - | 0.80,SI,0.91 | 0.81,0.7,0.91 |
| Tangent Shear Modulus [Pa] | γ=0.1% | SI,SI,M | S,SI,VS | SI,SI,S | M,SI,VS | S,SI,VS | S,SI,VS | VS,SI,VS | VS,SI,VS | - | 0.85,SI,0.91 |
| | γ=15% | SI,SI,M | M,SI,S | SI,S,S | M,SI,VS | VS,S,VS | S,M,VS | VS,S,VS | VS,S,VS | VS,SI,VS | - |

## 4    Discussion

In general, the results of our DSL experiments show good agreement with the early findings reported in the literature. Liu and Bilston [23] performed oscillation experiments at 37ºC on bovine liver tissue under a torsional shear strain of 0.11% and for the frequency range of 0.006-20 Hz. The shear storage and loss moduli of the bovine liver were varied between $G_S$=1000-7000 Pa and $G_L$=300-1000 Pa, respectively. Klatt et al. [33] conducted rheological experiments with bovine liver samples in the frequency range of 2.5-62.5 Hz frequency and at 0.3% torsional shear strain and 1ºC temperature. The storage and loss moduli were varied between $G_S$=1000-3000 Pa and $G_L$=400-1270 Pa, respectively. Nicolle et al. [28] performed rheological experiments on porcine liver samples at 37ºC in the frequency range of 0.1-4 Hz under a strain rate of 0.0151 s$^{-1}$, 0.133 s$^{-1}$ and 0.7 s$^{-1}$ and for the post-mortem period of 0-24 h. The storage and loss moduli of the samples were measured as $G_S$=800-1100 Pa and $G_L$=150-300 Pa, respectively. Tan et al. [24] investigated the effect of the pre-strain on the viscoelastic behavior of the liver tissue at 2-6 h post mortem time. The authors conducted amplitude sweep experiments at a frequency of 1 Hz, for shear strain value varying between 0.005-2%. The storage and loss moduli of liver tissue were reported to vary between $G_S$=350-450 Pa and $G_L$=70-90 Pa, respectively for the pre-strain value in the normal direction varying between 1-20%. Wex et al. [27] performed strain (1 Hz, 0.0001-1% strain), stress (1 Hz, 0.1-100 Pa), and frequency sweeps (0.1-10 Hz, 0.001% shear strain) and relaxation experiments (1 Hz, 1 Pa, 600 s) on porcine liver using a rheometer.

The magnitude of the complex shear modulus of porcine liver was reported to vary between |G*|=636-668 Pa under strain sweep for the post-mortem period of 1 h and as |G*|=443-538 Pa for the post-mortem period of 20-27 h. The magnitude of the complex shear modulus was reported to vary between |G*|=564-1004 Pa and |G*|=470-578 Pa under frequency sweep for the post-mortem

period of 1 h and 20-27 h, respectively. Valtorta and Mazza [34] used a torsional resonator to characterize the complex shear modulus of bovine liver (few hours after harvesting) for the frequency range of 1–10 kHz at the ambient temperature. The results of the in-vitro experiments suggested that the magnitude of the complex shear modulus of the liver varies between |G*|=5000 - 20000 Pa. Sapin-de Brosses et al. [35] measured the magnitude of complex shear modulus of bovine liver as |G*|=3400 ± 500 Pa by Supersonic shear imaging technique for the frequency range of 150–200 Hz at a temperature of 25ºC and for post-mortem period of 0-48 h. Riek et al. [36] used MRE technique to examine the dynamic mechanical properties of bovine liver tissue at a temperature of 17-19ºC for the frequency range of 100-800 Hz and post-mortem period of 0-2 h. The authors reported the storage and loss moduli of bovine liver tissue as $G_S$=1420-4910 Pa and $G_L$=740-1620 Pa, respectively. Kruse et al. [37] measured the magnitude of the complex shear modulus of the porcine liver at 37ºC as |G*|=3000 Pa at 100 Hz and as |G*|=5000 Pa at 300 Hz by using the MRE method. Saraf et al. [38] used the Kolsky bar technique and estimated the tangent shear modulus of the human liver under dynamic loading as G =37000-340000 Pa for the frequency range of 280 to 1800 Hz.

There are also studies in the literature reporting the dynamic elastic modulus of animal and human livers (Note that the linear elastic and shear moduli are related through $E = 2G(1 + v)$ for homogenous and isotropic materials. If the Poisson's ratio is taken as $v = 0.5$, then this relation reduces to $E = 3G$). DeWall et al. [39] used dynamic compression testing to quantify the viscoelastic properties of 16 human liver samples for a frequency range of 1—30 Hz under a compression strain of 2%, a pre-compression strain of 1-6%, and for a post-mortem period of 0-2 h. The elastic storage and loss moduli of the healthy human liver were measured as $E_S$=2000-6000 Pa and $E_L$=500-2000 Pa, respectively. Kiss et al. [22] applied cyclic stimuli (0.1 Hz to 400 Hz) to the canine liver tissue in the normal direction within 72 h of post-mortem time at 21°C and estimated the storage and loss modulus as $E_S$=4000-10000 Pa and $E_L$=800-10000 Pa, respectively. Ocal et al. [40] performed impact hammer experiments on bovine liver for the frequency range of 0-100 Hz at the room temperature and for post-mortem period of 1-48 h. They measured the elastic storage and loss moduli of the bovine liver as $E_S$=5000-80000 Pa and $E_L$=1000-13000 Pa, respectively.

There are several reasons for the discrepancy between the findings of the above studies. Obviously, the differences in the measurement devices, measurement methods, and the measured subjects (human, porcine, bovine, canine) contribute significantly to this discrepancy. Also, there are differences in the frequency of stimulation and several studies in the literature have already shown that the storage and loss moduli of the liver tissue are frequency-dependent. Another reason for the discrepancy could be the effect of temperature. The storage and loss moduli of the liver tissue have been shown to change with temperature as well. Furthermore, in our experiments, we observed the effect of pre-stress in the normal direction on the torsional shear stress measured by the rheometer (note that a certain amount of pre-stress in normal direction is necessary to hold the tissue sample between the parallel plates of the rheometer). While the effect of normal stress on rheological shear measurements has been mentioned briefly in a few studies and most authors report the magnitude of the normal force applied to the samples, we are not aware of any systematic study investigating the effect of this normal force on the torsional shear stress measurements. This topic needs a further investigation. Also, in our experiments, the tissue samples were prepared from parenchyma of each liver after removing its Glisson's capsule. We note that the Glisson's capsule is about three times stiffer than parenchyma and removing Glisson's capsule may lead to underestimation of the mechanical properties by 2-3 folds as pointed in Hollenstein et al (2006) [41].

In this study, we show that the preservation solution and period significantly affect the material properties of liver tissue. For example, we found that the change in the storage and loss moduli of the samples stored in HTK solution is far more less (about 1.1 folds, p <0.001) than those of the samples stored in UW solution (about 4 folds, p <0.001) at 53$^{rd}$ h. A similar observation can be made for the linear shear modulus of the samples. At small torsional shear strain of 0.1%, the tangent shear moduli of the samples preserved for 53 h in Ringer, UW, and HTK solutions are 2.38, 2.24, and 1.17 folds higher than those preserved in the same solutions for 5 h, respectively. We argue that the samples preserved in HTK solution have lower storage and loss moduli compared to those preserved in Ringer and UW solutions since the HTK is a low-viscosity solution [7]. Our results also show that the storage and loss moduli of liver tissue increase as it spends more time in preservation. Garo et al. [42], Kerdok et al. [43], Ocal et al. [40], and Yarpuzlu et al. [44] also observed that liver tissue becomes stiffer and more viscous as the post-mortem time increases. On the other hand, Wex et al. [27] have recently reported that the complex shear modulus of liver tissue decreases with increasing post-mortem times.

In addition to investigating mechanical properties of bovine liver, we also examined its histological properties as a function of preservation solution and period by various staining kits. The results suggest that the number of apoptotic cells, connective tissue, and sinusoidal dilatation increase while the glycogen level deceases as a function of preservation period. We observed less number of apoptotic cells in the samples stored in UW solution and less connective tissue (the main component of the ECM) accumulation in the samples preserved in HTK solution. These findings are compatible with the pharmacology of the preservation solutions. UW is known to be an intracellular type preservation solution and effective in protecting the integrity of the cell structure. On the other hand, HTK is an extracellular type preservation solution and effective in preserving the integrity of the ECM structure [7]. Also, our results are in agreement with the findings reported in the literature. Straatsburg et al. [13] analyzed the apoptosis in cold-preserved rat livers stored in HTK, UW, and Celsior solutions under a light microscope. The authors reported that the median number of apoptotic cells stored in UW solution is lower than those stored in HTK solution. Natori et al. [45] examined the level of apoptosis during cold preservation of liver in UW solution. The authors observed that the number of apoptotic cells have increased 6-folds after 24 h preservation. Abrahamse et al. [46] studied cell necrosis and apoptosis after 24 h and 48 h preservation of porcine liver in UW, HTK and Celsior solutions. The authors reported that UW and Celsior solutions are superior to HTK solution in terms of hepatocyte (liver cell) preservation. Lately, Budzinski et al. [47] assessed the cell apoptosis in porcine livers preserved for 12 h in UW and HTK solutions. For an apoptosis scale ranging from 0 to 3+, the authors reported the level of apoptotic bodies as +3 in both UW and HTK solutions. However, the accumulated connective tissue was higher for the samples preserved in UW solution. Puhl et al. [48] reported that the sinusoidal dilation is higher in rat liver cells preserved in HTK solution than those preserved in UW solution. Jain et al. [49] investigated the flow dynamics of sinusoidal perfusion during 24 h hypothermic machine perfusion of human livers preserved in UW solution. The authors observed an increase in sinusoidal dilatation as a function of preservation period due to perfusion damage. Furthermore, in our histological measurements, the glycogen level decreases as a function of increasing preservation period. This result suggests that certain percentage of hepatocytes could not synthesize and store glycogen since the deposited glycogen was consumed for the survival. In our study, the glycogen level was the highest in the samples stored in UW solution and lowest in the samples preserved in Ringer solution. Corps et al. [50] analyzed the adenine triphosphate and metabolites of the perfused rat livers preserved in HTK and UW solutions. The authors reported that livers perfused with UW solution do not consume glycogen as much as HTK-perfused livers due to the presence of alpha-ketoglutarate in HTK solution.

In this study, we also investigated the relation between the mechanical and histological properties of liver tissue as a function of preservation period and solution. The previous studies investigating the relation between the mechanical and histological properties of the liver tissue have mainly focused on the correlation between the stiffness of the diseased liver tissue at a certain frequency and its fibrosis score. Ozcan et al. [51], Mazza et al. [52], Sandrin et al. [53], Yeh et al. [54], Manduca et al. [55], Huwart et al. [56], Mori et al. [57], and Yarpuzlu et al. [44] have already observed a positive correlation between the accumulated connective tissue and an increase in liver stiffness. We performed a more in-depth study and investigated the correlation between histological properties (number of apoptotic cells, amount of connective tissue accumulation, degree of sinusoidal dilatation, and level of glycogen) and mechanical properties (storage and loss moduli, the tangent shear modulus) as a function of both preservation solution and period. Our results show that the apoptotic cell count, which is directly related to the graft's survival, is correlated with the mechanical properties of the liver tissue stored in UW solution. The depletion of the glycogen depositions, which is an early indicator for future cell deaths, is correlated with the mechanical properties of the liver tissue stored in Ringer and UW solutions. Also, the accumulated connective tissue, which is used to determine the stage of liver fibrosis, is solely correlated with the mechanical properties of the liver tissue stored in Ringer and UW solutions. Furthermore, the sinusoidal dilatation is moderately correlated with the dynamic mechanical properties, and strongly correlated with the quasi-static mechanical properties of the samples stored in UW solution. The results of our analyses demonstrated that the correlation between the inspected properties is solution-dependent. For instance, for the liver samples preserved in UW solution, both the histological and the mechanical properties change with respect to the preservation period. Therefore, the correlations between the histological and the mechanical properties are strong. However, when the liver tissue is preserved in HTK solution, the mechanical properties are preserved well and do not alter much, but the histological properties (except the connective tissue) alter significantly as a function of preservation period. Hence, the correlations between the mechanical and histological properties are mostly weak. In summary, these results suggest that the preservation solution must be taken into consideration while developing a diagnostic technique for the examination of the transplant suitability based on the correlation between the material and pathological properties.

Finally, we suggest an upper limit for the preservation period based on the statistical analyses performed on the mechanical and histological properties. For each preservation solution, we examined the statistical differences between the measurements performed at different preservation periods. The results reveal that UW solution preserves the histological properties of the liver tissue effectively for 29 h considering the apoptotic cell death and glycogen storage, 17 h considering the amount of the connective tissue and 53 h considering the sinusoidal dilatation. HTK solution preserves the histological properties of the liver tissue about 11 h considering the level of apoptosis and glycogen deposition, 53 h considering the connective tissue and 17 h considering the amount of sinusoidal dilatation. Also, it is seen that Ringer solution preserves the histological properties of the liver tissue about 11 h considering the apoptotic cell death and glycogen storage, 17 h considering the connective tissue, and 29 h considering the amount of sinusoidal dilatation. Also, based on the statistical analyses, the mechanical properties of the liver tissue are most effectively preserved in UW solution in the short run (about 29 h considering the storage and loss moduli and about 11 h considering the tangent shear modulus). In contrast, the mechanical properties of liver tissue are most successfully preserved in HTK solution in the long run (between 29-53 h considering the storage, loss, and tangent shear moduli). Moreover, these results also suggest that Ringer solution is the least effective solution for preserving the mechanical properties of the liver tissue (11 h considering the dynamic and quasi-static mechanical properties). To sum up, our results suggest that if the preservation time is expected to be shorter than 11 h, any of the three solutions (Ringer, HTK or UW) can be used safely. If the preservation period is expected to be shorter than 29 h, the use of UW solution is suggested. However, if the preservation period is expected to be longer than 29 h (but shorter than 53 h), HTK solution is a better option owing to its superiority in preserving the mechanical properties of the liver samples. The last observation was also supported by the small change in the connective tissue accumulation of the same samples preserved in HTK solution. In general, our results are in good agreement with the results of the earlier studies. Several studies have reported that there are no major differences between the solutions when the preservation period is short ([2]: within 10 h; [11]: within 15 h; and [13]: within 16 h). However, if the preservation period is longer, UW solution has shown to be superior to the other preservation solutions ([13]: after 16 h; [14]: between 18-42 h; [15]: after 24 h; [16]: after 24 h; [17]: after 24 h; [19]: after 24 h; and [18]: after 24 h).

## 5  Conclusion

We measured the mechanical properties of the bovine liver tissue stored in Ringer, HTK and UW solutions as function of preservation period using an oscillatory shear rheometer. We found that the storage, loss, and tangent shear moduli of liver tissue increase as it spends more time in the preservation solution. We observed that the largest increase in mechanical properties occur in the liver samples preserved in Ringer solution, while the smallest change occur in those stored in HTK solution. Also, the ratio between the loss and storage modulus (phase angle) increase for the bovine livers preserved in Ringer and UW solutions, while it stays almost constant for the liver samples preserved in HTK solution. In addition, we examined the histological properties of the same tissue samples as a function of preservation solution and period. We found that the number of apoptotic cells, the amount of connective tissue, and the amount of sinusoidal dilatation increase while the amount of glycogen deposition deceases as a function of increasing preservation period. These results provide an insight into the mechanism of tissue injury during cold ischemic preservation period. Since the blood is not supplied to the liver vessels during the cold ischemic preservation, the apoptosis is triggered, the connective tissue is produced as a control mechanism against the apoptosis, the sinusoids are dilated due to the venous flow impairment, and the glycogen supplies are depleted due to the energy consumption.

We also investigated the relation between the mechanical and histological properties of the liver tissue. Our statistical analyses show that the preservation solution and period have significant effects on the material and histological properties of the liver tissue. The earlier studies have only investigated the histological properties of the liver tissue to investigate the preservation solutions and period. In our study, we analyzed both the mechanical and histological properties of the liver tissue simultaneously to investigate the extent of safe preservation in Ringer, HTK and UW solutions. Based on our statistical analyses, we conclude that the bovine liver tissue is preserved well in Ringer, HTK and UW solutions up to 11 h. After then, UW solution provides the best preservation up to 29 h. If a preservation period longer than 29 h is necessary, HTK solution is a better alternative since the material properties of the liver samples stored in UW solution changed drastically after 29 h. It is important to note that although bovine and human liver tissues are similar to each other, they are not exactly the same. Hence, the histological and mechanical state of the liver of transplant patients should be also investigated to check the transferability of bovine results to human.

## Acknowledgement

The Scientific and Technological Research Council of Turkey (TUBITAK) supported this work under contract MAG-110M649 and the student fellowship program BIDEB-2211.